\documentstyle[11pt,paspconf]{article}
\def\PsfigVersion{1.10}
\def\setDriver{\DvipsDriver} 
\ifx\undefined\psfig\else \fi
\let\LaTeXAtSign=\@
\let\@=\relax
\edef\psfigRestoreAt{\catcode`\@=\number\catcode`@\relax}
\catcode`\@=11\relax
\newwrite\@unused
\def\ps@typeout#1{{\let\protect\string\immediate\write\@unused{#1}}}

\def\DvipsDriver{
	\ps@typeout{psfig/tex \PsfigVersion -dvips}
\def\PsfigSpecials{\DvipsSpecials} 	\def\ps@dir{/}
\def\ps@predir{} }
\def\OzTeXDriver{
	\ps@typeout{psfig/tex \PsfigVersion -oztex}
	\def\PsfigSpecials{\OzTeXSpecials}
	\def\ps@dir{:}
	\def\ps@predir{:}
	\catcode`\^^J=5
}

\def\figurepath{./:}

\def\DoPaths#1{\expandafter\EachPath#1\stoplist}
\def\leer{}
\def\EachPath#1:#2\stoplist{
  \ExistsFile{#1}{\SearchedFile}
  \ifx#2\leer
  \else
    \expandafter\EachPath#2\stoplist
  \fi}
\def\ps@dir{/}
\def\ExistsFile#1#2{%
   \openin1=\ps@predir#1\ps@dir#2
   \ifeof1
       \closein1
   \else
       \closein1
        \ifx\ps@founddir\leer
           \edef\ps@founddir{#1}
        \fi
   \fi}
\def\get@dir#1{%
  \def\ps@founddir{}
  \def\SearchedFile{#1}
  \DoPaths\figurepath
}

\def\@nnil{\@nil}
\def\@empty{}
\def\@psdonoop#1\@@#2#3{}
\def\@psdo#1:=#2\do#3{\edef\@psdotmp{#2}\ifx\@psdotmp\@empty \else
    \expandafter\@psdoloop#2,\@nil,\@nil\@@#1{#3}\fi}
\def\@psdoloop#1,#2,#3\@@#4#5{\def#4{#1}\ifx #4\@nnil \else
       #5\def#4{#2}\ifx #4\@nnil \else#5\@ipsdoloop #3\@@#4{#5}\fi\fi}
\def\@ipsdoloop#1,#2\@@#3#4{\def#3{#1}\ifx #3\@nnil 
       \let\@nextwhile=\@psdonoop \else
      #4\relax\let\@nextwhile=\@ipsdoloop\fi\@nextwhile#2\@@#3{#4}}
\def\@tpsdo#1:=#2\do#3{\xdef\@psdotmp{#2}\ifx\@psdotmp\@empty \else
    \@tpsdoloop#2\@nil\@nil\@@#1{#3}\fi}
\def\@tpsdoloop#1#2\@@#3#4{\def#3{#1}\ifx #3\@nnil 
       \let\@nextwhile=\@psdonoop \else
      #4\relax\let\@nextwhile=\@tpsdoloop\fi\@nextwhile#2\@@#3{#4}}
\ifx\undefined\fbox
\newdimen\fboxrule
\newdimen\fboxsep
\newdimen\ps@tempdima
\newbox\ps@tempboxa
\long\def\fbox#1{\leavevmode\setbox\ps@tempboxa\hbox{#1}\ps@tempdima\fboxrule
    \advance\ps@tempdima \fboxsep \advance\ps@tempdima \dp\ps@tempboxa
   \hbox{\lower \ps@tempdima\hbox
  {\vbox{\hrule height \fboxrule
          \hbox{\vrule width \fboxrule \hskip\fboxsep
          \vbox{\vskip\fboxsep \box\ps@tempboxa\vskip\fboxsep}\hskip 
                 \fboxsep\vrule width \fboxrule}
                 \hrule height \fboxrule}}}}
\fi
\newread\ps@stream
\newif\ifnot@eof       
\newif\if@noisy        
\newif\if@atend        
\newif\if@psfile       
{\catcode`\%=12\global\gdef\epsf@start{
\def\epsf@PS{PS}
\def\epsf@getbb#1{%
\openin\ps@stream=\ps@predir#1
\ifeof\ps@stream\ps@typeout{Error, File #1 not found}\else
   {\not@eoftrue \chardef\other=12
    \def\do##1{\catcode`##1=\other}\dospecials \catcode`\ =10
    \loop
       \if@psfile
	  \read\ps@stream to \epsf@fileline
       \else{
	  \obeyspaces
          \read\ps@stream to \epsf@tmp\global\let\epsf@fileline\epsf@tmp}
       \fi
       \ifeof\ps@stream\not@eoffalse\else
       \if@psfile\else
       \expandafter\epsf@test\epsf@fileline:. \\%
       \fi
          \expandafter\epsf@aux\epsf@fileline:. \\%
       \fi
   \ifnot@eof\repeat
   }\closein\ps@stream\fi}%
\long\def\epsf@test#1#2#3:#4\\{\def\epsf@testit{#1#2}
			\ifx\epsf@testit\epsf@start\else
\ps@typeout{Warning! File does not start with `\epsf@start'.  It may not be a PostScript file.}
			\fi
			\@psfiletrue} 
{\catcode`\%=12\global\let\epsf@percent=
\long\def\epsf@aux#1#2:#3\\{\ifx#1\epsf@percent
   \def\epsf@testit{#2}\ifx\epsf@testit\epsf@bblit
	\@atendfalse
        \epsf@atend #3 . \\%
	\if@atend	
	   \if@verbose{
		\ps@typeout{psfig: found `(atend)'; continuing search}
	   }\fi
        \else
        \epsf@grab #3 . . . \\%
        \not@eoffalse
        \global\no@bbfalse
        \fi
   \fi\fi}%
\def\epsf@grab #1 #2 #3 #4 #5\\{%
   \global\def\epsf@llx{#1}\ifx\epsf@llx\empty
      \epsf@grab #2 #3 #4 #5 .\\\else
   \global\def\epsf@lly{#2}%
   \global\def\epsf@urx{#3}\global\def\epsf@ury{#4}\fi}%
\def\epsf@atendlit{(atend)} 
\def\epsf@atend #1 #2 #3\\{%
   \def\epsf@tmp{#1}\ifx\epsf@tmp\empty
      \epsf@atend #2 #3 .\\\else
   \ifx\epsf@tmp\epsf@atendlit\@atendtrue\fi\fi}

\chardef\psletter = 11 
\chardef\other = 12

\newif \ifdebug 
\newif\ifc@mpute 
\c@mputetrue 

\let\then = \relax
\def\r@dian{pt }
\let\r@dians = \r@dian
\let\dimensionless@nit = \r@dian
\let\dimensionless@nits = \dimensionless@nit
\def\internal@nit{sp }
\let\internal@nits = \internal@nit
\newif\ifstillc@nverging
\def \Mess@ge #1{\ifdebug \then \message {#1} \fi}

{ 
	\catcode `\@ = \psletter
	\gdef \nodimen {\expandafter \n@dimen \the \dimen}
	\gdef \term #1 #2 #3%
	       {\edef \t@ {\the #1}
		\edef \t@@ {\expandafter \n@dimen \the #2\r@dian}%
		\t@rm {\t@} {\t@@} {#3}%
	       }
	\gdef \t@rm #1 #2 #3%
	       {{%
		\count 0 = 0
		\dimen 0 = 1 \dimensionless@nit
		\dimen 2 = #2\relax
		\Mess@ge {Calculating term #1 of \nodimen 2}%
		\loop
		\ifnum	\count 0 < #1
		\then	\advance \count 0 by 1
			\Mess@ge {Iteration \the \count 0 \space}%
			\Multiply \dimen 0 by {\dimen 2}%
			\Mess@ge {After multiplication, term = \nodimen 0}%
			\Divide \dimen 0 by {\count 0}%
			\Mess@ge {After division, term = \nodimen 0}%
		\repeat
		\Mess@ge {Final value for term #1 of 
				\nodimen 2 \space is \nodimen 0}%
		\xdef \Term {#3 = \nodimen 0 \r@dians}%
		\aftergroup \Term
	       }}
	\catcode `\p = \other
	\catcode `\t = \other
	\gdef \n@dimen #1pt{#1} 
}

\def \Divide #1by #2{\divide #1 by #2} 

\def \Multiply #1by #2
       {{
	\count 0 = #1\relax
	\count 2 = #2\relax
	\count 4 = 65536
	\Mess@ge {Before scaling, count 0 = \the \count 0 \space and
			count 2 = \the \count 2}%
	\ifnum	\count 0 > 32767 
	\then	\divide \count 0 by 4
		\divide \count 4 by 4
	\else	\ifnum	\count 0 < -32767
		\then	\divide \count 0 by 4
			\divide \count 4 by 4
		\else
		\fi
	\fi
	\ifnum	\count 2 > 32767 
	\then	\divide \count 2 by 4
		\divide \count 4 by 4
	\else	\ifnum	\count 2 < -32767
		\then	\divide \count 2 by 4
			\divide \count 4 by 4
		\else
		\fi
	\fi
	\multiply \count 0 by \count 2
	\divide \count 0 by \count 4
	\xdef \product {#1 = \the \count 0 \internal@nits}%
	\aftergroup \product
       }}

\def\r@duce{\ifdim\dimen0 > 90\r@dian \then   
		\multiply\dimen0 by -1
		\advance\dimen0 by 180\r@dian
		\r@duce
	    \else \ifdim\dimen0 < -90\r@dian \then  
		\advance\dimen0 by 360\r@dian
		\r@duce
		\fi
	    \fi}

\def\Sine#1%
       {{%
	\dimen 0 = #1 \r@dian
	\r@duce
	\ifdim\dimen0 = -90\r@dian \then
	   \dimen4 = -1\r@dian
	   \c@mputefalse
	\fi
	\ifdim\dimen0 = 90\r@dian \then
	   \dimen4 = 1\r@dian
	   \c@mputefalse
	\fi
	\ifdim\dimen0 = 0\r@dian \then
	   \dimen4 = 0\r@dian
	   \c@mputefalse
	\fi
	\ifc@mpute \then
		\divide\dimen0 by 180
		\dimen0=3.141592654\dimen0
		\dimen 2 = 3.1415926535897963\r@dian 
		\divide\dimen 2 by 2 
		\Mess@ge {Sin: calculating Sin of \nodimen 0}%
		\count 0 = 1 
		\dimen 2 = 1 \r@dian 
		\dimen 4 = 0 \r@dian 
		\loop
			\ifnum	\dimen 2 = 0 
			\then	\stillc@nvergingfalse 
			\else	\stillc@nvergingtrue
			\fi
			\ifstillc@nverging 
			\then	\term {\count 0} {\dimen 0} {\dimen 2}%
				\advance \count 0 by 2
				\count 2 = \count 0
				\divide \count 2 by 2
				\ifodd	\count 2 
				\then	\advance \dimen 4 by \dimen 2
				\else	\advance \dimen 4 by -\dimen 2
				\fi
		\repeat
	\fi		
			\xdef \sine {\nodimen 4}%
       }}

\def\Cosine#1{\ifx\sine\UnDefined\edef\Savesine{\relax}\else
		             \edef\Savesine{\sine}\fi
	{\dimen0=#1\r@dian\advance\dimen0 by 90\r@dian
	 \Sine{\nodimen 0}
	 \xdef\cosine{\sine}
	 \xdef\sine{\Savesine}}}	      

\def\psdraft{
	\def\@psdraft{0}
}
\def\psfull{
	\def\@psdraft{100}
}

\psfull

\newif\if@scalefirst
\def\psscalefirst{\@scalefirsttrue}
\def\psrotatefirst{\@scalefirstfalse}
\psrotatefirst

\newif\if@draftbox
\def\psnodraftbox{
	\@draftboxfalse
}
\def\psdraftbox{
	\@draftboxtrue
}
\@draftboxtrue

\newif\if@prologfile
\newif\if@postlogfile
\def\pssilent{
	\@noisyfalse
}
\def\psnoisy{
	\@noisytrue
}
\psnoisy
\newif\if@bbllx
\newif\if@bblly
\newif\if@bburx
\newif\if@bbury
\newif\if@height
\newif\if@width
\newif\if@rheight
\newif\if@rwidth
\newif\if@angle
\newif\if@clip
\newif\if@verbose
\def\@p@@sclip#1{\@cliptrue}
\newif\if@decmpr
\def\@p@@sfigure#1{\def\@p@sfile{null}\def\@p@sbbfile{null}\@decmprfalse
   \openin1=\ps@predir#1
   \ifeof1
	\closein1
	\get@dir{#1}
	\ifx\ps@founddir\leer
		\openin1=\ps@predir#1.bb
		\ifeof1
			\closein1
			\get@dir{#1.bb}
			\ifx\ps@founddir\leer
				\ps@typeout{Can't find #1 in \figurepath}
			\else
				\@decmprtrue
				\def\@p@sfile{\ps@founddir\ps@dir#1}
				\def\@p@sbbfile{\ps@founddir\ps@dir#1.bb}
			\fi
		\else
			\closein1
			\@decmprtrue
			\def\@p@sfile{#1}
			\def\@p@sbbfile{#1.bb}
		\fi
	\else
		\def\@p@sfile{\ps@founddir\ps@dir#1}
		\def\@p@sbbfile{\ps@founddir\ps@dir#1}
	\fi
   \else
	\closein1
	\def\@p@sfile{#1}
	\def\@p@sbbfile{#1}
   \fi
}
\def\@p@@sfile#1{\@p@@sfigure{#1}}
\def\@p@@sbbllx#1{
		\@bbllxtrue
		\dimen100=#1
		\edef\@p@sbbllx{\number\dimen100}
}
\def\@p@@sbblly#1{
		\@bbllytrue
		\dimen100=#1
		\edef\@p@sbblly{\number\dimen100}
}
\def\@p@@sbburx#1{
		\@bburxtrue
		\dimen100=#1
		\edef\@p@sbburx{\number\dimen100}
}
\def\@p@@sbbury#1{
		\@bburytrue
		\dimen100=#1
		\edef\@p@sbbury{\number\dimen100}
}
\def\@p@@sheight#1{
		\@heighttrue
		\dimen100=#1
   		\edef\@p@sheight{\number\dimen100}
}
\def\@p@@swidth#1{
		\@widthtrue
		\dimen100=#1
		\edef\@p@swidth{\number\dimen100}
}
\def\@p@@srheight#1{
		\@rheighttrue
		\dimen100=#1
		\edef\@p@srheight{\number\dimen100}
}
\def\@p@@srwidth#1{
		\@rwidthtrue
		\dimen100=#1
		\edef\@p@srwidth{\number\dimen100}
}
\def\@p@@sangle#1{
		\@angletrue
		\edef\@p@sangle{#1} 
}
\def\@p@@ssilent#1{ 
		\@verbosefalse
}
\def\@p@@sprolog#1{\@prologfiletrue\def\@prologfileval{#1}}
\def\@p@@spostlog#1{\@postlogfiletrue\def\@postlogfileval{#1}}
\def\@cs@name#1{\csname #1\endcsname}
\def\@setparms#1=#2,{\@cs@name{@p@@s#1}{#2}}
\def\ps@init@parms{
		\@bbllxfalse \@bbllyfalse
		\@bburxfalse \@bburyfalse
		\@heightfalse \@widthfalse
		\@rheightfalse \@rwidthfalse
		\def\@p@sbbllx{}\def\@p@sbblly{}
		\def\@p@sbburx{}\def\@p@sbbury{}
		\def\@p@sheight{}\def\@p@swidth{}
		\def\@p@srheight{}\def\@p@srwidth{}
		\def\@p@sangle{0}
		\def\@p@sfile{} \def\@p@sbbfile{}
		\def\@p@scost{10}
		\def\@sc{}
		\@prologfilefalse
		\@postlogfilefalse
		\@clipfalse
		\if@noisy
			\@verbosetrue
		\else
			\@verbosefalse
		\fi
}
\def\parse@ps@parms#1{
	 	\@psdo\@psfiga:=#1\do
		   {\expandafter\@setparms\@psfiga,}}
\newif\ifno@bb
\def\bb@missing{
	\if@verbose{
		\ps@typeout{psfig: searching \@p@sbbfile \space  for bounding box}
	}\fi
	\no@bbtrue
	\epsf@getbb{\@p@sbbfile}
        \ifno@bb \else \bb@cull\epsf@llx\epsf@lly\epsf@urx\epsf@ury\fi
}	
\def\bb@cull#1#2#3#4{
	\dimen100=#1 bp\edef\@p@sbbllx{\number\dimen100}
	\dimen100=#2 bp\edef\@p@sbblly{\number\dimen100}
	\dimen100=#3 bp\edef\@p@sbburx{\number\dimen100}
	\dimen100=#4 bp\edef\@p@sbbury{\number\dimen100}
	\no@bbfalse
}
\newdimen\p@intvaluex
\newdimen\p@intvaluey
\def\rotate@#1#2{{\dimen0=#1 sp\dimen1=#2 sp
		  \global\p@intvaluex=\cosine\dimen0
		  \dimen3=\sine\dimen1
		  \global\advance\p@intvaluex by -\dimen3
		  \global\p@intvaluey=\sine\dimen0
		  \dimen3=\cosine\dimen1
		  \global\advance\p@intvaluey by \dimen3
		  }}
\def\compute@bb{
		\no@bbfalse
		\if@bbllx \else \no@bbtrue \fi
		\if@bblly \else \no@bbtrue \fi
		\if@bburx \else \no@bbtrue \fi
		\if@bbury \else \no@bbtrue \fi
		\ifno@bb \bb@missing \fi
		\ifno@bb \ps@typeout{FATAL ERROR: no bb supplied or found}
			\no-bb-error
		\fi
		\count203=\@p@sbburx
		\count204=\@p@sbbury
		\advance\count203 by -\@p@sbbllx
		\advance\count204 by -\@p@sbblly
		\edef\ps@bbw{\number\count203}
		\edef\ps@bbh{\number\count204}
		\if@angle 
			\Sine{\@p@sangle}\Cosine{\@p@sangle}
	        	{\dimen100=\maxdimen\xdef\r@p@sbbllx{\number\dimen100}
					    \xdef\r@p@sbblly{\number\dimen100}
			                    \xdef\r@p@sbburx{-\number\dimen100}
					    \xdef\r@p@sbbury{-\number\dimen100}}
                        \def\minmaxtest{
			   \ifnum\number\p@intvaluex<\r@p@sbbllx
			      \xdef\r@p@sbbllx{\number\p@intvaluex}\fi
			   \ifnum\number\p@intvaluex>\r@p@sbburx
			      \xdef\r@p@sbburx{\number\p@intvaluex}\fi
			   \ifnum\number\p@intvaluey<\r@p@sbblly
			      \xdef\r@p@sbblly{\number\p@intvaluey}\fi
			   \ifnum\number\p@intvaluey>\r@p@sbbury
			      \xdef\r@p@sbbury{\number\p@intvaluey}\fi
			   }
			\rotate@{\@p@sbbllx}{\@p@sbblly}
			\minmaxtest
			\rotate@{\@p@sbbllx}{\@p@sbbury}
			\minmaxtest
			\rotate@{\@p@sbburx}{\@p@sbblly}
			\minmaxtest
			\rotate@{\@p@sbburx}{\@p@sbbury}
			\minmaxtest
			\edef\@p@sbbllx{\r@p@sbbllx}\edef\@p@sbblly{\r@p@sbblly}
			\edef\@p@sbburx{\r@p@sbburx}\edef\@p@sbbury{\r@p@sbbury}
		\fi
		\count203=\@p@sbburx
		\count204=\@p@sbbury
		\advance\count203 by -\@p@sbbllx
		\advance\count204 by -\@p@sbblly
		\edef\@bbw{\number\count203}
		\edef\@bbh{\number\count204}
}
\def\in@hundreds#1#2#3{\count240=#2 \count241=#3
		     \count100=\count240	
		     \divide\count100 by \count241
		     \count101=\count100
		     \multiply\count101 by \count241
		     \advance\count240 by -\count101
		     \multiply\count240 by 10
		     \count101=\count240	
		     \divide\count101 by \count241
		     \count102=\count101
		     \multiply\count102 by \count241
		     \advance\count240 by -\count102
		     \multiply\count240 by 10
		     \count102=\count240	
		     \divide\count102 by \count241
		     \count200=#1\count205=0
		     \count201=\count200
			\multiply\count201 by \count100
		 	\advance\count205 by \count201
		     \count201=\count200
			\divide\count201 by 10
			\multiply\count201 by \count101
			\advance\count205 by \count201
		     \count201=\count200
			\divide\count201 by 100
			\multiply\count201 by \count102
			\advance\count205 by \count201
		     \edef\@result{\number\count205}
}
\def\compute@wfromh{
		\in@hundreds{\@p@sheight}{\@bbw}{\@bbh}
		\edef\@p@swidth{\@result}
}
\def\compute@hfromw{
	        \in@hundreds{\@p@swidth}{\@bbh}{\@bbw}
		\edef\@p@sheight{\@result}
}
\def\compute@handw{
		\if@height 
			\if@width
			\else
				\compute@wfromh
			\fi
		\else 
			\if@width
				\compute@hfromw
			\else
				\edef\@p@sheight{\@bbh}
				\edef\@p@swidth{\@bbw}
			\fi
		\fi
}
\def\compute@resv{
		\if@rheight \else \edef\@p@srheight{\@p@sheight} \fi
		\if@rwidth \else \edef\@p@srwidth{\@p@swidth} \fi
}
\def\compute@sizes{
	\compute@bb
	\if@scalefirst\if@angle
	\if@width
	   \in@hundreds{\@p@swidth}{\@bbw}{\ps@bbw}
	   \edef\@p@swidth{\@result}
	\fi
	\if@height
	   \in@hundreds{\@p@sheight}{\@bbh}{\ps@bbh}
	   \edef\@p@sheight{\@result}
	\fi
	\fi\fi
	\compute@handw
	\compute@resv}
\def\OzTeXSpecials{
	\special{empty.ps /@isp {true} def}
	\special{empty.ps \@p@swidth \space \@p@sheight \space
			\@p@sbbllx \space \@p@sbblly \space
			\@p@sbburx \space \@p@sbbury \space
			startTexFig \space }
	\if@clip{
		\if@verbose{
			\ps@typeout{(clip)}
		}\fi
		\special{empty.ps doclip \space }
	}\fi
	\if@angle{
		\if@verbose{
			\ps@typeout{(rotate)}
		}\fi
		\special {empty.ps \@p@sangle \space rotate \space} 
	}\fi
	\if@prologfile
	    \special{\@prologfileval \space } \fi
	\if@decmpr{
		\if@verbose{
			\ps@typeout{psfig: Compression not available
			in OzTeX version \space }
		}\fi
	}\else{
		\if@verbose{
			\ps@typeout{psfig: including \@p@sfile \space }
		}\fi
		\special{epsf=\ps@predir\@p@sfile \space }
	}\fi
	\if@postlogfile
	    \special{\@postlogfileval \space } \fi
	\special{empty.ps /@isp {false} def}
}
\def\DvipsSpecials{
	\special{ps::[begin] 	\@p@swidth \space \@p@sheight \space
			\@p@sbbllx \space \@p@sbblly \space
			\@p@sbburx \space \@p@sbbury \space
			startTexFig \space }
	\if@clip{
		\if@verbose{
			\ps@typeout{(clip)}
		}\fi
		\special{ps:: doclip \space }
	}\fi
	\if@angle
		\if@verbose{
			\ps@typeout{(clip)}
		}\fi
		\special {ps:: \@p@sangle \space rotate \space} 
	\fi
	\if@prologfile
	    \special{ps: plotfile \@prologfileval \space } \fi
	\if@decmpr{
		\if@verbose{
			\ps@typeout{psfig: including \@p@sfile.Z \space }
		}\fi
		\special{ps: plotfile "`zcat \@p@sfile.Z" \space }
	}\else{
		\if@verbose{
			\ps@typeout{psfig: including \@p@sfile \space }
		}\fi
		\special{ps: plotfile \@p@sfile \space }
	}\fi
	\if@postlogfile
	    \special{ps: plotfile \@postlogfileval \space } \fi
	\special{ps::[end] endTexFig \space }
}
\def\psfig#1{\vbox {
	\ps@init@parms
	\parse@ps@parms{#1}
	\compute@sizes
	\ifnum\@p@scost<\@psdraft{
		\PsfigSpecials 
		\vbox to \@p@srheight sp{
			\hbox to \@p@srwidth sp{
				\hss
			}
		\vss
		}
	}\else{
		\if@draftbox{		
			\hbox{\fbox{\vbox to \@p@srheight sp{
			\vss
			\hbox to \@p@srwidth sp{ \hss 
			 \hss }
			\vss
			}}}
		}\else{
			\vbox to \@p@srheight sp{
			\vss
			\hbox to \@p@srwidth sp{\hss}
			\vss
			}
		}\fi

	}\fi
}}
\psfigRestoreAt
\setDriver
\let\@=\LaTeXAtSign

\def\eg{{e.g.~}}
\def\lsim{\;\raise0.3ex\hbox{$<$\kern-0.75em\raise-1.1ex\hbox{$\sim$}}\;}
\def\kmm{\mbox{\,km~s$^{-1}$ Mpc$^{-1}$}}

\begin{document}
\title{ The CASTLES gravitational lensing tool}
\author{E. E. Falco, C. S. Kochanek, J. Leh\'ar, B. A. McLeod, 
J. A. Mu\~noz} 
\affil{Harvard-Smithsonian Center for Astrophysics\\ 
           60 Garden St., Cambridge, MA 02138}
\author{C. D. Impey, C. Keeton, C. Y. Peng} 
\affil{Steward Observatory, University of Arizona\\
           Tucson, AZ 85721}
\author{H.-W. Rix}
\affil{Max-Planck Institut f\"ur Astrophysik, Heidelberg, Germany}
 
\begin{abstract}
We describe a series of new applications of gravitational lenses as
astrophysical and cosmological tools. Such applications are becoming
possible thanks to advances in the quality and quantity of
observations. CASTLES ({\bf C}fA-{\bf A}rizona-{\bf S}pace-{\bf
T}elescope-{\bf LE}ns-{\bf S}urvey)\footnote
{cfa-www.harvard.edu/castles} is an ongoing project that exploits the
sensitivity and resolution of the Hubble Space Telescope (HST) at
optical and infrared wavelengths to study the sample of over 50 known
gravitational lenses. The observational goal of CASTLES is a uniform
sample of multi-band images of all known galaxy-mass lens systems, to
derive precise photometry and astrometry for the lens galaxies, all
the known images, and any source or lens components that might have
escaped detection. With these measurements we are investigating: (1)
the properties of dust and of dark matter in lens galaxies out to
$z\sim1$; (2) the dark matter in lens galaxies and in their
environments; (3) the evolution of lens galaxies; and (4) the
cosmological model, for instance by refining constraints on the Hubble
constant $H_0$.

\end{abstract}


\vspace*{-0.5cm}
\section{Introduction}

There are now 54 known gravitational lens systems where a single,
usually early-type galaxy has been identified as the dominant
component.  Lens galaxies are unique, because they constitute the only
sample selected based on mass rather than luminosity.  The population
of known lenses is growing exponentially, with an e-folding time of
about 5 years.  The broad astrophysical utility of the lens sample
rests on obtaining accurate photometry, astrometry and redshifts for
the complete sample.  Since these lens systems consist of 2--4 source
images (AGNs, quasars, hosts), separated by $1\arcsec$--$2\arcsec$ and
centered on a foreground lens galaxy, quantitative studies at optical
to IR wavelengths require HST.

The CASTLES observational goals are to assemble a complete, uniform
high-resolution photometric sample of all known galaxy-mass lenses.
The sample yields precise photometry and astrometry for all the
components in each lens system, and for source (e.g., host galaxies
for lensed quasars) or lens components (e.g., galaxies near the main
lens galaxy) that had previously escaped detection.  With our analysis
of these measurements, we are pursuing various astrophysical and
cosmological goals: (1) measuring the extinction in the lens galaxies
using the photometric properties of multiple images; (2) obtaining
redshift estimates for all the lens galaxies without measured
spectroscopic redshifts; (3) improving lens models to thus refine
estimates of $H_0$; and (4) investigating the tidal field environments
of the lens galaxies and their role in lensing.

For large population surveys of lenses like CASTLES we require a
uniform data quality. Unfortunately, many archival lens observations
were not dithered, had inadequate exposure times, or used
idiosyncratic filter choices, all of which can make their utilization
in population surveys impossible.  Archival images are included in our
data sample only when their quality is sufficient for our analysis.
The CASTLES filters (V and I on WFPC2; H on NIC2) match the usual
choices for studies of other galaxies at comparable redshifts.
CASTLES began in HST Cycle 7, and is continuing in Cycles 8 and 9.  Of
the 54 known lenses, 60\%, 70\% and 80\% now have V, I and H-band HST
images that are useful for our scientific goals. Our source for
targets has been a variety of ground-based discoveries of promising
candidates.  Time is being reserved in each of our proposals for
systems that are discovered in the course of each HST Cycle. A
significant number of targets became available in Cycles 7 and 8, for
example.  Thus, we are able to observe a number of targets of
opportunity as ground-based observers provide us with new, promising
candidates.  All our data are available to the community as they
arrive, as we waive all proprietary restrictions.

The WFPC2 and NICMOS point-spread functions (PSF) are complex, and can
vary spatially and temporally.  We used TinyTim v4.4 (Krist \& Hook
1997) model PSFs for WFPC2 and NIC1 data, while for models of the NIC2
data we used a set of 13 stellar images (McLeod, Rieke \&
Storrie-Lombardi 1999).  By comparing the PSFs to stellar images, we
found that the residuals were less than 2\% of the peak intensity of
the stars.  We used our library of empirical PSFs to find the best fit
for each lens system.  We homogeneously reduce and model all CASTLES
and archival data, using software we have developed for these
purposes, e.g. the NICRED reduction package\footnote{available for
download at cfa-www.harvard.edu/castles}.

In the remaining section, we present results that address several of
the CASTLES goals. We describe first several ``missing'' lenses that
were first found in our observations.  We then present new detections
of lensed host galaxies.  Next, we present results regarding
photometric lens redshifts and extinction, galaxy structure, the
properties of dark matter in lens galaxies, and finally the CASTLES
contribution to measurements of $H_0$.

\section{CASTLES results}

\noindent{\bf Missing Lenses:} As expected, CASTLES optical and IR
observations are detecting the majority of ``missing'' lens galaxies,
even when bright quasar images and small separations make this
difficult. For example, among the sample of 10 doubles in Leh\'ar et
al. (1999), we detected 9 lens galaxies, 3 of which are new detections
(see Fig. 1).  We failed only in the case of Q~1208+1011, due to
extreme contrast problems.  We also failed to detect lenses
among the wider-separation ($\geq 3\arcsec$) quasar pairs UM~425,
Q~1429--008, Q~1634+267 (Peng et al. 1999), MGC~2214+3550, and
Q~2345+007 (see below), whose extreme limits on the lens $M/L$ make
lensing improbable and point to the binary quasar interpretation
(Jackson et al. 1998; Kochanek, Falco \& Mu\~noz, 1999).

\noindent {\bf Host Galaxies:} With our CASTLES images, we have
detected lensed AGN host galaxies for 7 quasars (CTQ~414, 
MG~0414+0534, BRI~0952--0115,
Q~0957+561, HE1104--1805, PG~1115+080 and H~1413+117), 
and 5 radio galaxies (MG~0751\-+2716, MG~1131+0456,
B~1600+434, B~1608+656 and B~1938+666).  Lensing provides a unique
opportunity to study $z > 1$ host galaxies, because the hosts are
magnified to detectable angular sizes (see Fig. 2 for an example). We
find that $L_{\rm host}\lsim L_\ast$, i.e., that even the most
luminous quasars (\eg PG1115+080) need not reside in particularly
luminous hosts (see Rix et al., these proceedings).  Our new H-band
image of Q~0957+561 shows 2 images of the quasar host galaxy (Fig. 2);
their geometry rules out the existing lens models for the system
(Keeton et al.  1999). In combination with existing WFPC2 V-band data
(Bernstein \& Fischer 1999) and with new STIS imaging (Bernstein et
al., these proceedings), our data yield useful new models for
Q~0957+561 (Keeton et al. 1999, in preparation).  For the QSO
pair Q~2345+007 (Fig. 2), the host morphology 
proves the system is a binary quasar and not a ``dark'' lens, because
only one of the quasars is embedded in a host (Fig. 2).
 
\begin{figure}
\centerline{
  \psfig{figure=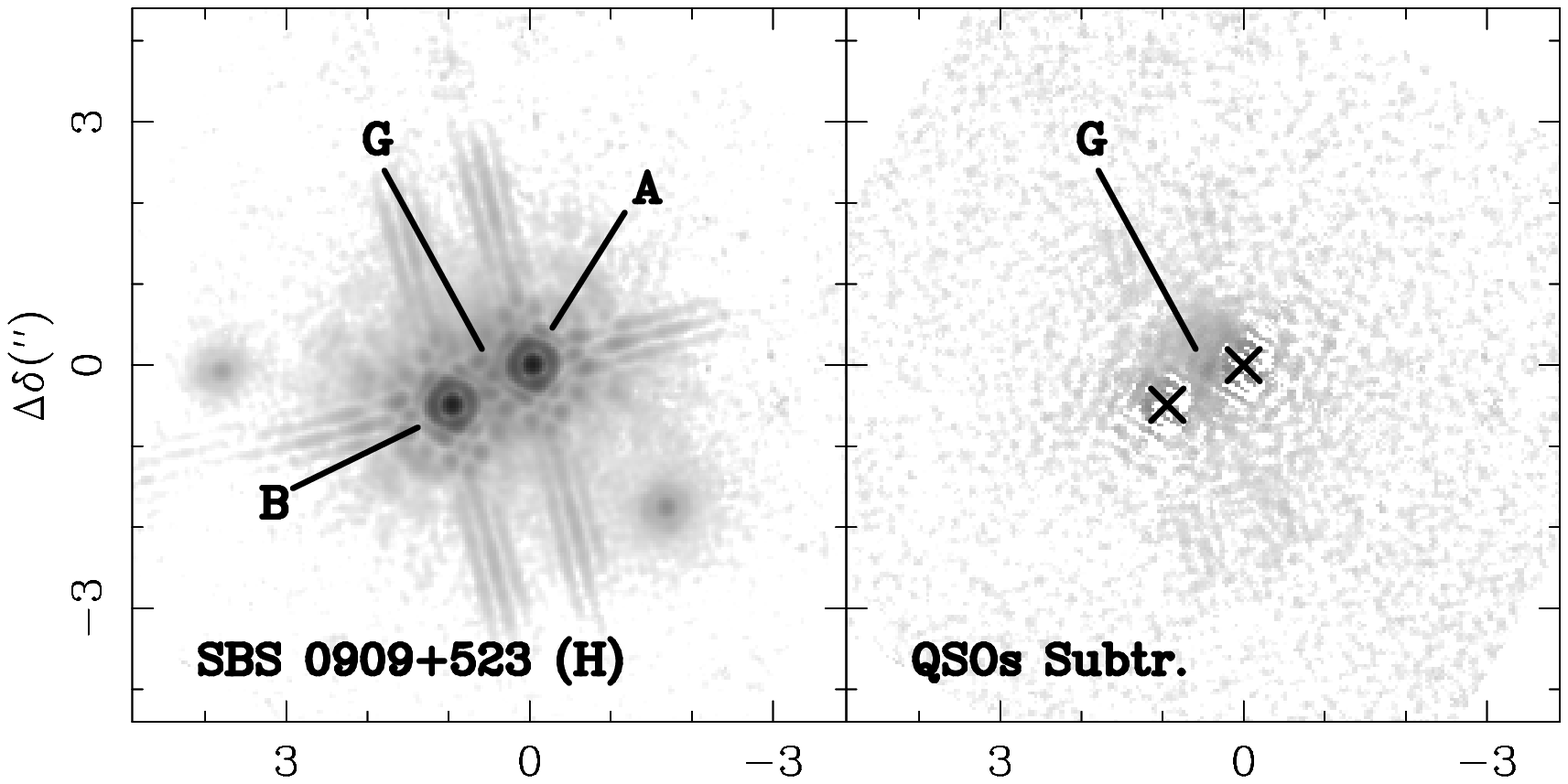,width=4.8in,angle=0.}
}
\vspace*{-1.cm}
\centerline{
  \psfig{figure=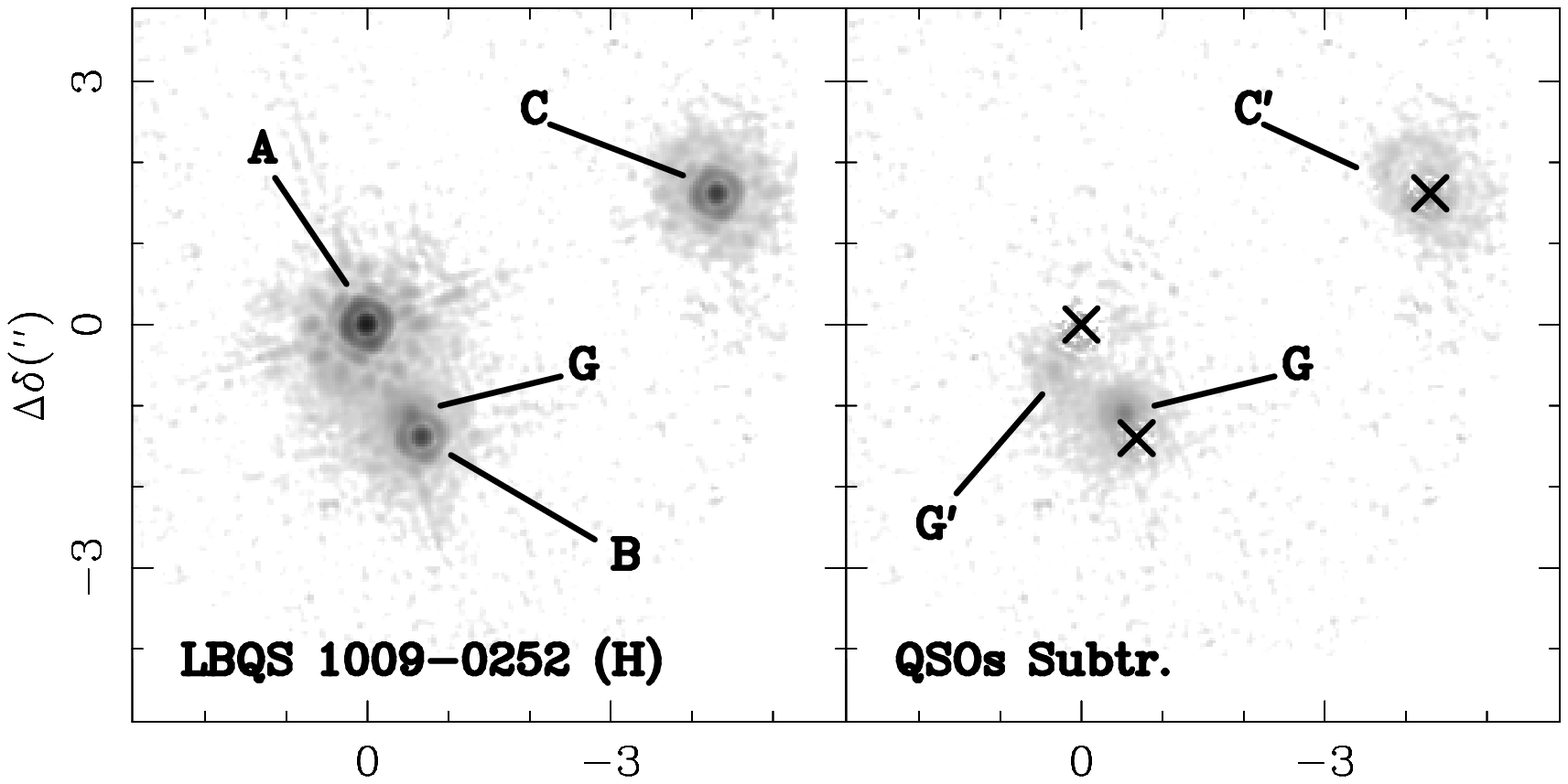,width=4.8in,angle=0.}
}
\vspace*{-1.cm}
\centerline{
  \psfig{figure=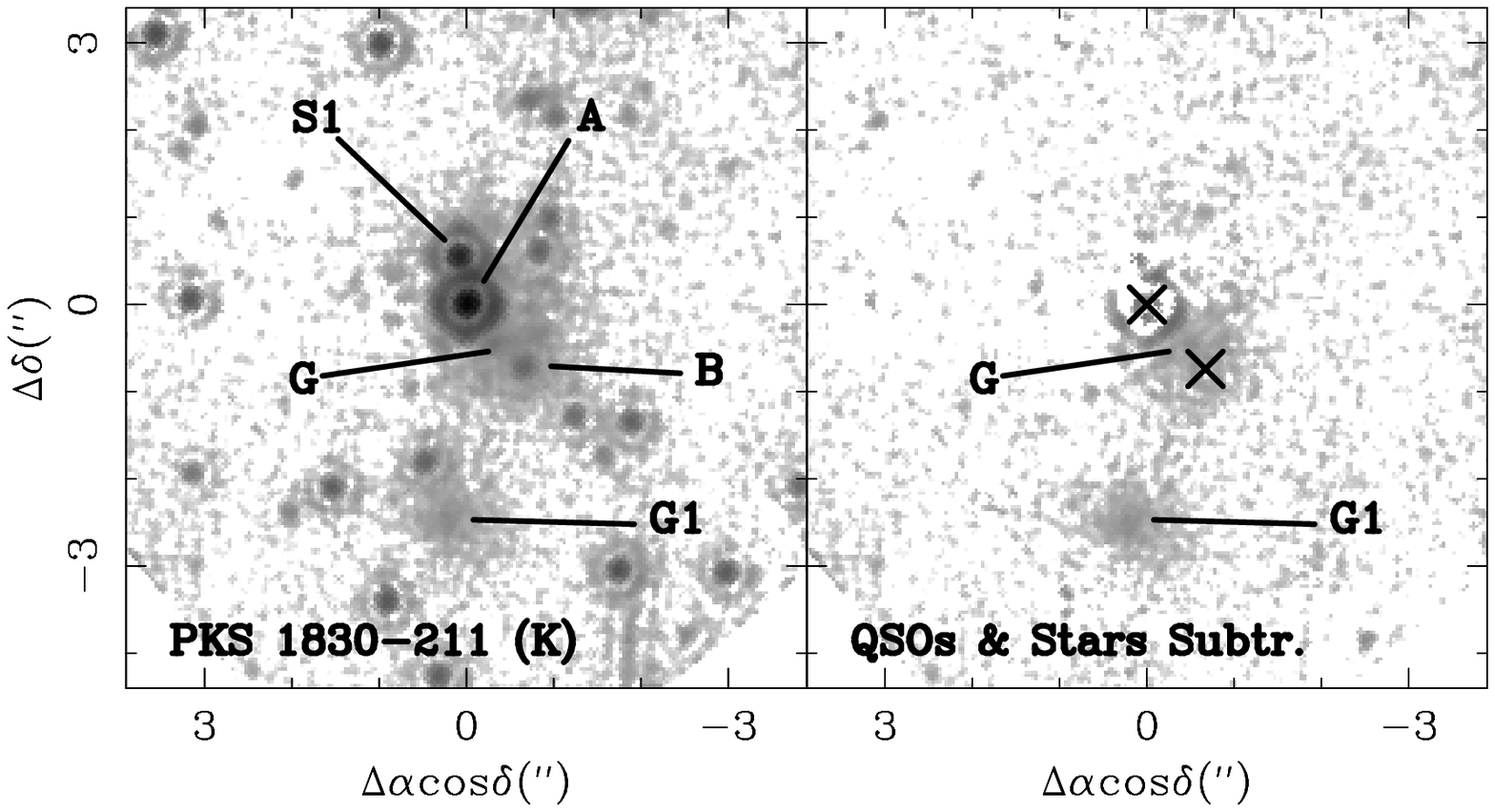,width=4.8in,angle=0.}
}
\vspace*{-0.5cm}
\caption{Missing lenses: top to bottom, the left panels show CASTLES
NICMOS images of the systems SBS~0909+523 and LBQS~1009--0252 in the H
band and of PKS~1830--211 in the K band (thus, the B image remains 
detectable, see Leh\'ar et al. 1999).  The scale is the same
for all the panels.  The right panels show the results of subtracting
the lensed quasar images (and the stars near PKS~1830--211).  }
\end{figure}

\begin{figure}
\centerline{
  \psfig{figure=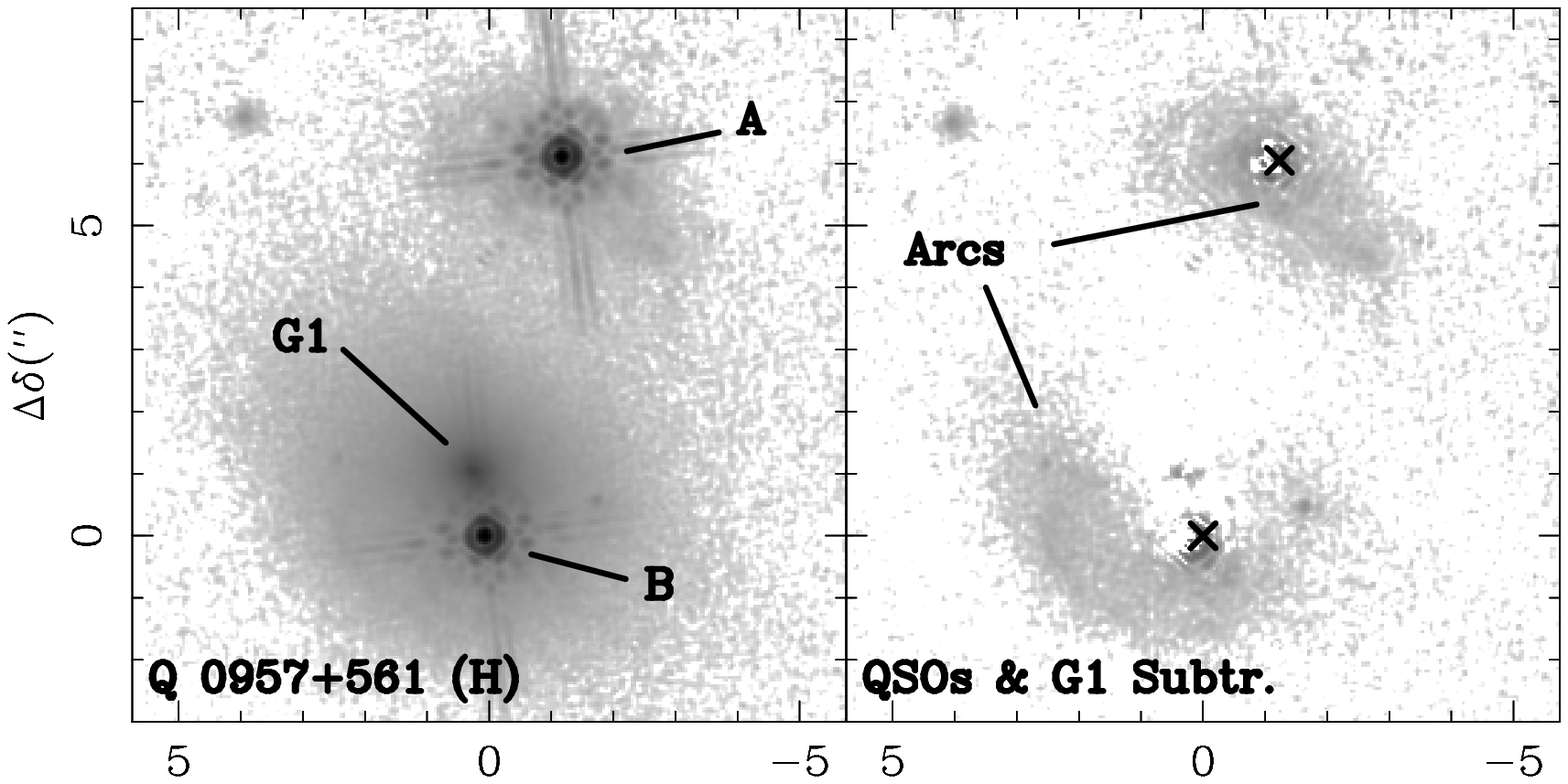,width=4.8in,angle=0.}
}
\vspace*{-1.cm}
\centerline{
  \psfig{figure=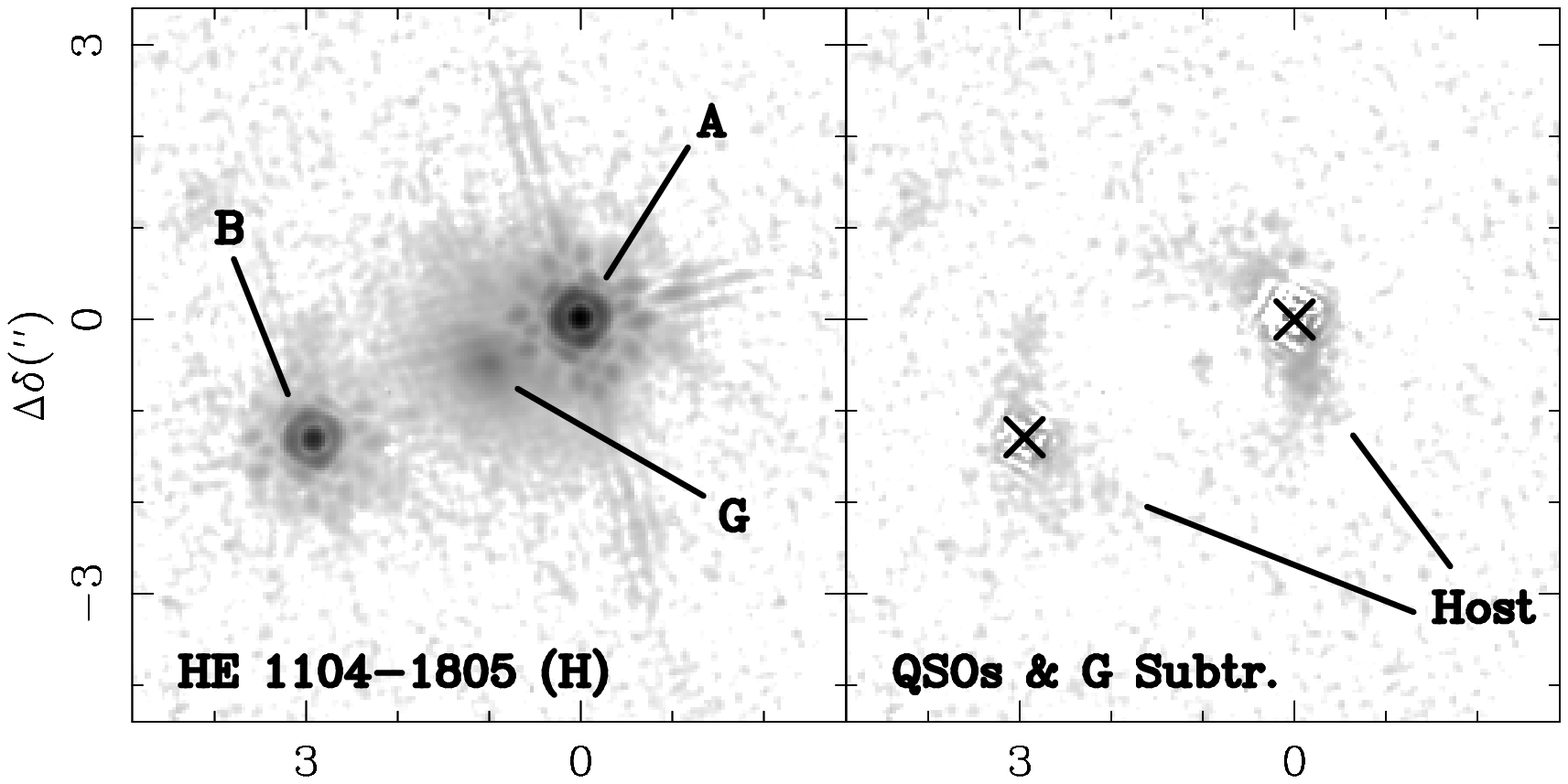,width=4.8in,angle=0.}
}
\vspace*{-1.cm}
\centerline{
  \psfig{figure=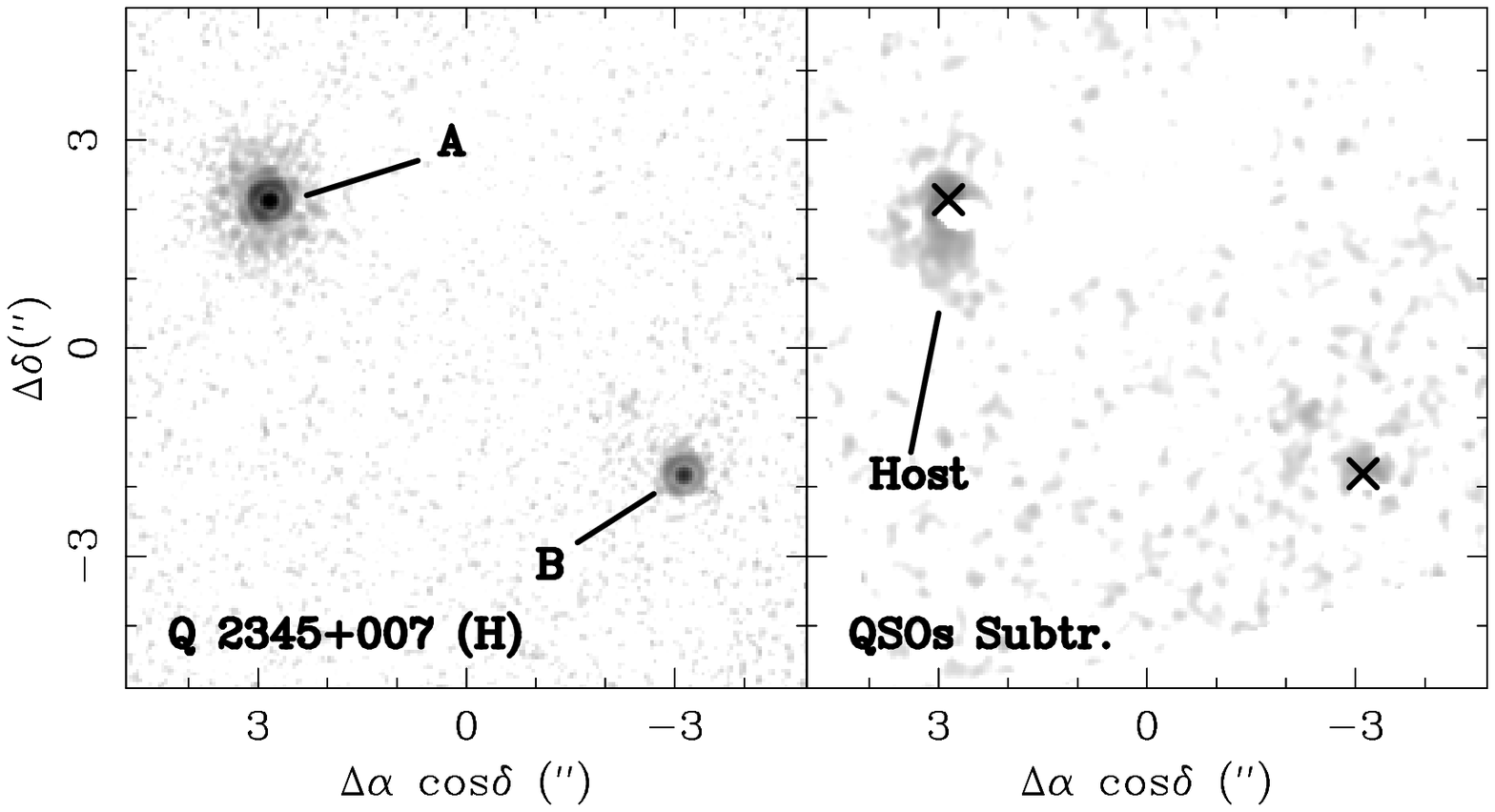,width=4.8in,angle=0.}
}
\vspace*{-0.5cm}
\caption{Host galaxies: top to bottom, the left panels show CASTLES
NICMOS H band images of Q~0957+561, HE~1104--1805 and Q~2345+007. The
right panels show the results of subtracting the lensed quasar images
and the lens galaxy. Note that the scales necessarily vary from image
to image.  }
\end{figure}

\noindent{\bf Galaxy Structure:} CASTLES observations provide detailed
constraints on lens galaxy structure, through shape and profile
fitting.  We find that most lens galaxies have de~Vaucouleurs
profiles, as well as shapes and colors consistent with passively
evolving early-type galaxies; e.g. Q~0142--100, BRI~0952--0115,
Q~1017--207, B~1030+071, HE~1104--1805 (Leh\'ar et al. 1999); MG~1131+0456
(Kochanek et al. 1999a) and PG~1115+080 (Impey et al. 1998).  B~0218+357
and PKS~1830--211 (Leh\'ar et al. 1999) are spirals but present a more
complex photometric picture due to the high molecular gas column
densities and implied dust extinction. Another clear late-type is
B~1600+434, where the lens is edge-on and a dust lane is clearly
visible (Fig. 3). The photometric model in this case requires an
exponential disk and a De Vaucouleurs profile for the bulge.

\noindent{\bf Dark Matter:} We constrain and study the dark matter
distribution of lenses by comparing their luminosity distribution to
lens mass models based on our CASTLES astrometry and photometry.  We
used either a constant M/L lens model matched to our photometric fits
or a singular isothermal ellipsoid.  We fit each model in isolation
and then with an external shear to represent perturbations to the
model from nearby galaxies or potential perturbations along the ray
paths.  Without external tidal perturbations, the constant M/L models
usually could not fit the lens constraints, and the dark matter models
could only fit the lens constraints if misaligned relative to the
luminosity.  With the addition of a modest external shear, either
model could fit all the lenses because the two-image lenses provide so
few constraints on the models.  In general, the constant M/L and dark
matter models predict very different time delays for the two-image
lenses, and are hence distinguishable (see, e.g., Leh\'ar et al. 1999).

\begin{figure}
\centerline{\psfig{figure=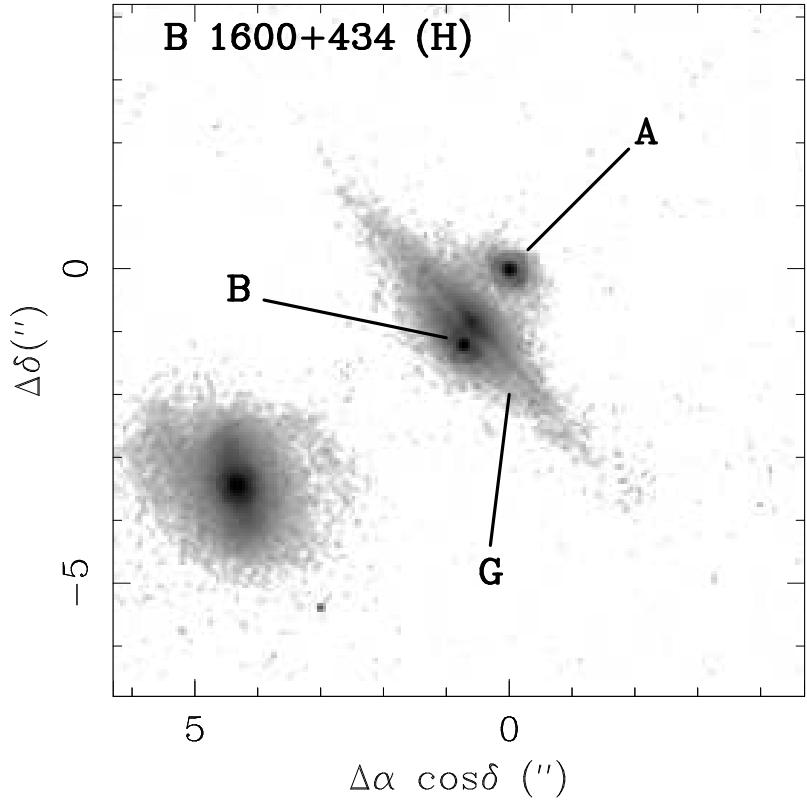,height=4.5in}}
\vspace*{-0.5cm}
\caption{CASTLES NIC2 H band image of B~1600+434. The 2 images A, B of
a radio galaxy are extended. The lens galaxy G is an edge-on
spiral. The nearby galaxy to the SE of the images presumably has the
same redshift as G; 
it is likely to be a significant contributor to the lensing. 
The dust lane is easily visible along the edge of the disk
of the lens galaxy G.}
\end{figure}

We also cataloged the galaxies found within $\sim 100h^{-1}$~kpc of
the lenses and estimated their local tidal environment based on the
luminosities of the neighboring galaxies.  If the galaxies have
extended dark matter halos, they produce significant shears at the
lens galaxy of $\gamma_T \sim 0.05$, while if they have a constant M/L
they are a negligible perturbation of $\gamma_T < 0.01$.  The dark
matter estimates for the tidal shear are comparable to the shears
necessary to find good lens models, but there is no clear correlation
between the shear from the lens model and the estimate from the local
environment.  A significant problem is that the strength of the shear
perturbations created by large-scale structure along the ray paths
(Bar-Kana 1996, Keeton et~al. 1997) is comparable to that from nearby
galaxies.

\noindent{\bf Lens Galaxy Extinction:} We determined differential
extinctions in 23 gravitational lens galaxies over the range $z_l =
0-1$ (Falco et al. 1999).  Only 7 of the 23 systems have spectral
differences consistent with no differential extinction (McLeod et al.,
these proceedings).  The extinction is patchy and shows no correlation
with impact parameter.  The directly measured extinction distributions
are consistent with the mean extinction estimated by comparing the
statistics of quasar and radio lens surveys.  For several systems we
can estimate the extinction law; the results range from $R_V=1.5\pm0.2$
for a $z_l=0.96$ elliptical, to $R_V=7.2\pm0.1$ for a $z_l=0.68$
spiral.  The dust can be used to estimate lens redshifts
with reasonable accuracy, although we sometimes find two degenerate
redshift solutions. We find agreement within $\Delta z_L\lsim 0.1$
with spectroscopic redshifts (Kochanek et al. 1999b).
 
\begin{figure}
\centerline{ \psfig{figure=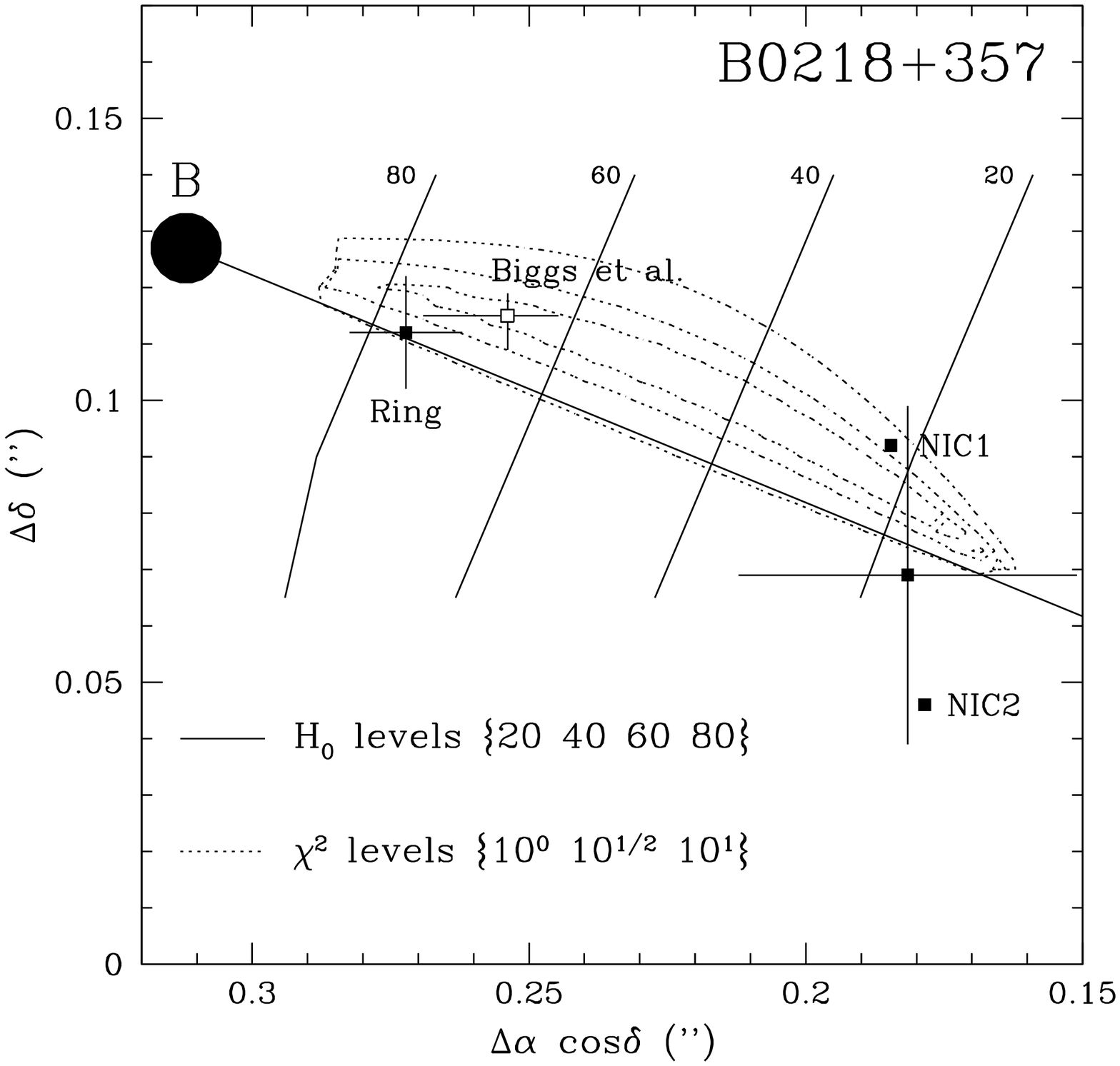,width=2.5in} 
             \psfig{figure=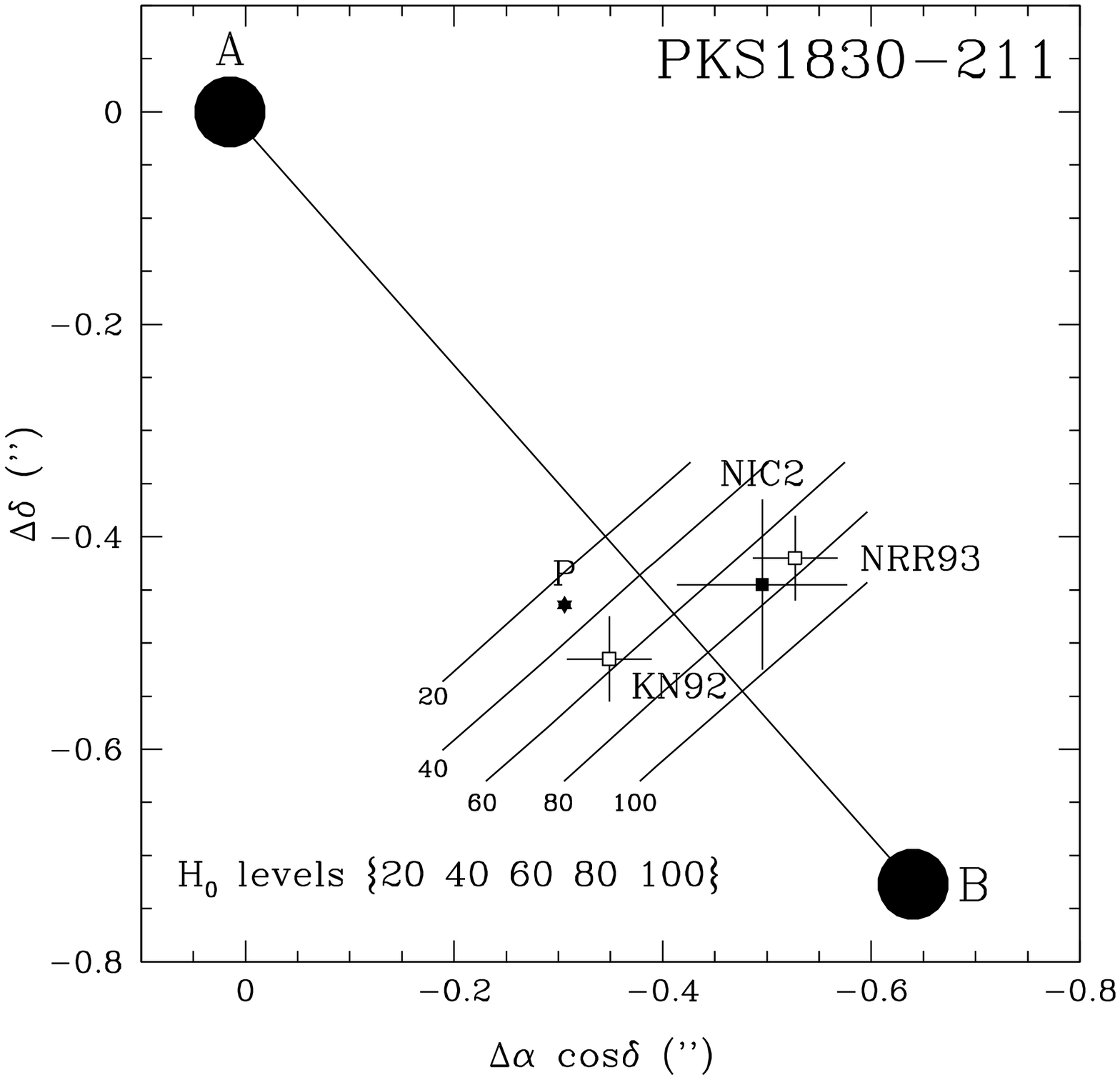,width=2.5in} }
\vspace*{-0.5cm}
\caption{Solid contours show the variation in $H_0$ with the position
of the lens given the measured time delays.  For B~0218+357 the
positions from the NICMOS observations are the filled points
NIC1 and NIC2; the point with large error bars is our estimate for
the lens position.  ``Ring'' marks the center of the
Einstein ring (Patnaik et~al. 1993; Biggs et~al. 1999). Filled
points for PKS~1830--211 mark the positions of the lens galaxy
(labeled NIC2).  Open points mark the lens positions in the models of
Kochanek \& Narayan (1992, KN92) and Nair et~al. (1993, NRR93). }
\end{figure}

\noindent{\bf The Hubble constant:} There are now 6 lens systems with
reliably measured time delays; 3 more show variability that
should yield measurements within the year (e.g., Burud et al., these
proceedings).  To translate the time delays into $H_0$ estimates,
accurate models for the mass distributions of the lenses are
essential.  Accurate models require precision astrometry and
additional constraints to minimize systematic uncertainties and thus
turn lenses into the best means of estimating $H_0$.

Among our observed targets, B~0218+357 (Biggs et~al.\ 1999) and
PKS~1830--211 (Lovell et~al.\ 1998) have measured time delays.
Unfortunately, with the short exposures necessary for a survey such as
CASTLES, we were unable to measure the positions of the lens galaxies
in these two systems with sufficient precision to accurately determine
$H_0$ (Fig. 4 illustrates the problem; see Leh\'ar et al 1999).  For
PKS~1830--211 only longer observations are necessary: our data
demonstrate that the uncertainties would be easily
reduced to acceptable levels given a higher signal-to-noise ratio
(SNR).  For B~0218+357, we are limited by the small component
separations rather than the SNR.  Better image sampling combined with
long exposures and concurrent PSF measurements should resolve the
problem.

\noindent{\bf Acknowledgements:} Support for the CASTLES project was
provided by NASA through grant numbers GO-7495 and GO-7887 from the
Space Telescope Science Institute which is operated by the Association
of Universities for Research in Astronomy, Inc. under NASA contract
NAS 5-26555.

\end{document}